\documentstyle[12pt]{article}  
\begin{document}  
\newtheorem{thm}{Theorem}
\newtheorem{cor}{Corollary}
\newtheorem{Def}{Definition}
\newtheorem{lem}{Lemma}
\begin{center}  
{\large \bf  
Quantum Mechanics and the Metrics of  General Relativity\\ 
Paul O'Hara\\ }  
\vspace{5mm}  
{\small\it  
Dept. of Mathematics, Northeastern Illinois University, 5500  
North St. Louis Avenue, Chicago, IL 60625-4699, USA. email:
pohara@neiu.edu \\}  
\end{center}  
%
%
%

\begin{abstract}
A one-to-one correspondence is established between 
linearized space-time metrics of general relativity and the wave equations
of quantum mechanics.  
Also, the key role of boundary conditions in distinguishing quantum mechanics
from classical mechanics, will emerge naturally from the procedure. 
Finally, we will find that the methodology will 
enable us to introduce not only test charges but also test masses by means of 
gauges. 

Pacs: 3.65, 4.60.
\end{abstract}

\section {Introduction}

There is a fundamental paradigm shift between general relativity 
and classical 
mechanics, characterized by the 
fact that in general relativity
the energy-momentum tensor is 
the effective cause of the space-time curvature,
while  in classical
physics, the structure of space-time is treated as an 
accidental
cause, serving only as a backdrop against which the laws of physics unfold.
This split in turn is inherited by quantum mechanics, which is usually
developed by changing classical (including special relativity) Hamiltonians into quantum wave equations. 
In this paper, we will try to remedy this situation by taking the metrics of 
general relativity as our starting point of quantum mechanics. We will 
associate wave equations in a natural way with those operators which are
duals of differential one-forms 
rather than with operators derived from a Hamiltonian, thus enabling the 
structure of space-time itself
to determine in a natural and unique way the wave
equations of quantum mechanics. Moreover, it is precisely the presence of gauge terms in the 
form of test masses and charge that 
permit new laws of physics to emerge independently within the context of the 
space-time structure in which they are embedded.
Throughout the paper, $({\cal M},g)$
\cite{haw} will denote a space-time pair,
where ``$\cal M$ is a connected four dimensional Hausdorff manifold''
and $g$ is a metric of signature -2 on $\cal M$.
At every point $p\in \cal M$ we can erect a local tetrad
$e_0(p), e_2(p), e_2(p), e_3(p)$
and a point $x$ will have coordinates $x=(x^0,\ x^1,\ x^2,\ x^3)=x^ae_a$ in
this tetrad coordinate system.\cite{oraf}
We will use Roman letters $a, b, c$
etc. to index coordinates with respect to a tetrad. In this regard, we will
refer to $x^a$ as the coordinates of $x$.
Also for spinors, we can write
$\psi =\psi^ie_i(p)$, where $\psi^i$ will represent the coordinates of the
spinor with respect to the tetrad at $p$. Also at $p$ we can establish a
tangent vector space $T_p(\cal M)$, with basis $\{\partial_1, \partial_2,
\partial_3, \partial_4\}$ and a dual 1-form space, denoted by $T^*_p$ with
basis $\{dx_1,dx_2,dx_3,dx_4\}$ at $p$, defined by 
\begin{eqnarray}
dx^{\mu}\partial_{\nu}\equiv
\partial_{\nu} x^{\mu}=\delta^{\mu}_{\nu}.
\end{eqnarray}
We refer to the basis
$\{dx^1, dx^2, dx^3, dx^4\}$ as ``the basis of one forms dual to the basis
$\{\partial_1, \partial_2, \partial_3, \partial_4\}$ 
of vectors at $p$."\cite{haw}
Finally, note that Greek letters
will be used to represent general coordinate systems and Einsteinian notation
will be used for summations.
\section {Metrics and the Dirac equation}

We begin with an intuitive and non-rigorous approach to our 
methodology by indicating two ways in which 
quantum mechanical wave equations can be obtained from the metrics of 
general relativity, 
without any explicit recourse to Lagrangians or Hamiltonians. We will then
combine the results of the two approaches into a mathematical theorem.
Later in the next section,
we will impose more rigorous constraints, which will enable us to identify the 
spinor formulation given here with the usual Hilbert Space formulation of 
quantum mechanics.

1) In a previous paper \cite{ohara} we have shown that the quantum-mechanical 
wave equations can be derived as the dual of the Dirac ``square-root'' of the
metric.
In other words, if
\begin{eqnarray} ds^2=g_{\mu \nu}dx^{\mu}dx^{\nu}=\eta_{ab}dx^adx^b 
\end{eqnarray}
where $a$ and $b$ refer to local tetrad coordinates and $\eta$ to a rigid 
Minkowski metric of signature -2, then associated with this 
metric and the vector $\vec ds$ is the scalar $ds$ and a matrix  
$\tilde{ds}\equiv\gamma_{a}dx^{a}$ respectively,
where $\{\gamma_a,\gamma_b\}=2\eta_{ab}$, with $\gamma_a$ transforming 
as a covariant vector under coordinate transformations.
Note also that $g_{\mu \nu}(x)=\eta_{ab}e^a_{\mu}(x)e^b_{\nu}(x)$
with $e^a_{\mu}(x)$ forming local tetrads\cite{oraf} at $x$.
Moreover, since $ds$ is an invariant scalar, and 
$\tilde{ds}^2=ds^2$ we can identify the ``eigenvalue'' $ds$ with 
the linear operator $\tilde{ds}$ by forming the spinor eigenvector
equation $\tilde{ds}\xi = ds\xi$ (see also Cartan p 106).
This is equivalent to associating the metric  
\begin{eqnarray} ds^2=g_{\mu \nu}dx^{\mu}dx^{\nu}=\eta_{ab}dx^adx^b 
\end{eqnarray}
with the spinor equation:
\begin{eqnarray}ds\xi=\gamma_{a}dx^{a}\xi.\end{eqnarray}
Note immediately that in agreement with the general theory of eigenvectors, if 
$\xi$ is a solution so also is $f(z_0, z_1, z_2, z_3)\xi$ where $f$ is any 
complex scalar valued function. Indeed, there is no reason why $f$ cannot be
an $L^2$ function, and correspond to a quantum-mechanical wave function.

As previously noted, corresponding to each tangent vector
$\frac{\partial}{\partial x^a}$, there exists a dual one-form $dx^a$.
In a similar way, the $\tilde{ds}$ matrix above can be seen as the dual
of the expression $\tilde{\partial}_s\equiv \gamma^a\frac{\partial}
{\partial x^a}$, where $\gamma^a$ is defined by the relationship
$\{\gamma^a,\gamma_b\}=2\delta^a_b$ and the dual map defined by
\begin{equation}\left<\tilde{ds},\tilde{\partial}_s\right>\equiv
\frac{1}{{\rm Tr}(dx^i\partial_j)}
\gamma_a\gamma^bdx^a\frac{\partial}{\partial x^b}\equiv
\frac{1}{\delta^i_i}
\gamma_a\gamma^b\frac{\partial x^a}{\partial x^b}=1,\end{equation}
remains invariant.
Moreover, if we let $s$ describe the length of a particle's trajectory
along a curve $(x^0(s), x^1(s), x^2(s), x^3(s))\in ({\cal M}, g)$ 
then $s$ can be regarded as an independent parameter with an
associated  1-form $ds$, which is the dual of the tangent vector
$\partial_s$. Note that in terms of the basis vectors for $T_p(\cal M)$ and
$T^*_p(\cal M)$ we can write $\partial_s =\frac{\partial x^a}{\partial s}\partial_a$
and $ds=\frac{\partial s}{\partial x^a}dx^a$. It also follows from this and equation (1) that its dual map is given by
$ds.\partial_s=\frac{\partial s}{\partial x^i}\frac{\partial x^i}{\partial s}=1$.
Putting these two results together allows us to consider equation (4)
as the dual of the equation:
\begin{eqnarray} \frac {\partial \psi}{\partial s}=\gamma^{a}
\frac{\partial \psi}
{\partial x^a}, \end{eqnarray}
where $\frac{\partial }{\partial s}$
refers to differentiation along a curve parametrized by $s$.
We will refer to (6) as a (generalized)
Dirac equation and will show later on how it relates to the usual form of this
equation. At times, too, we shall refer loosely to it as a
``dual wave-equation.''

Now consider the motion of a test particle of mass $m$ along a timelike 
geodesic. Let $p^a=m(dx^a/d\tau)$, where $\tau$ is the 
proper time (i.e $ds=cd\tau$). Then 
\begin{eqnarray} ds^2 = \eta_{ab}dx^adx^b \qquad \mbox {is equivalent to}
\qquad (mc)^2=\eta_{ab}p^ap^b.
\end{eqnarray}
This can be expressed in spinor notation by
\begin{eqnarray} ds\xi = \gamma_adx^a\xi \qquad \mbox {which is equivalent to}
\qquad \gamma^ap_a\xi(p)=mc\xi(p).
\end{eqnarray}
Indeed, if equation (6) is subjected to the constraints of equation (8), 
as it should be for motion along the timelike geodesic, we find that
$\psi =\psi^i(\int^{x(s)}_{x_0} p_adx^a)e_i$ is a solution of equation (6),
provided the integration is taken along the curve $s=c\tau$ and 
$\xi(p)=\frac{d\psi^i(p)}{d\tau}e_i$ (cf Theorem 1 below). 
It is also worth noting that if all of $\psi^i$ are equal
then $\psi =f(x)u$ where $u$ is a spinor independent of $x$, and $f(x)$
is a function. In this particular case the Dirac equation takes on the
form
\begin{eqnarray}(\tilde{\partial}_s f)u = 
\frac{\partial f}{\partial s}u.\end{eqnarray}
Moreover, in terms of a 4-dimensional (complex) Euclidean space $E^4$, this 
equation can be directly related to the expression\cite{car} 
\begin{equation}df=\vec {ds}.\nabla f= ds\frac{\partial f}{\partial s},
\end{equation}
by noting that $\tilde{\partial}_sf$ is the matrix form of the vector
$\nabla f$. Also, if $\vec ds.\nabla f$ is 
invariant with respect to both rotations and reflections, then
the associated spinor
equation can be immediately written in a covariant manner 
in a natural way.  It is
sufficient to note that for the Lorentz spinor transformation $D(\Lambda(x))$
applied to the Dirac equation (9), we get
\begin{eqnarray}D(\Lambda(x))(\tilde{\partial}_s f)u &=& 
D(\Lambda(x))\frac{\partial f}{\partial s}u\\
\Rightarrow\qquad D(\Lambda(x))(\tilde{\partial}_s f)D^{-1}(\Lambda(x))D(\Lambda(x))u
&=&\frac{\partial f(s))}{\partial s}D(\Lambda(x))u\\
\Rightarrow\qquad \tilde{\partial f}_{s^{\prime}}u^{\prime}(x^{\prime})
&=&\frac{\partial f(x)}{\partial s}u^{\prime}(x^{\prime}),
\end{eqnarray}\newline
which expresses the covariance. Note $\tilde{\partial}_{s^{\prime}}\equiv
\gamma^a\frac{\partial}{\partial x^{\prime}_a}$.
\newline

2) Another approach to the above formalism is to introduce test 
particles by means of a gauge term. Specifically, let $({\cal M}, g)$ be a
pseudo-Riemannian manifold, 
with metric tensor $g$ determined by Einstein's field equations. 
Consider a massless test particle introduced into the field, then by the
principle of equivalence, we can choose a local tetrad $\{dx_1, dx_2, dx_3,
dx_4\}$, such that the massless test particle travels along a null geodesic
given by 
\begin{eqnarray*} 0& = & dx^adx_a,
\end{eqnarray*}
which in terms of a spinor basis can be written as
\begin{eqnarray}
\gamma^adx_a \xi &=& 0.
\end{eqnarray}
Next define the wave equation corresponding to the metric by taking the dual
of the 1-form space: 
\begin{eqnarray}
\gamma^a\partial_a \psi &=& 0, 
\end{eqnarray}
which can be interpreted as the wave equation of a massless particle.

We now introduce a test particle of mass $m$ by means of a minimal principle 
\cite{oraf}, by adopting the
same technique that is usually used to introduce test charges into a field. 
In other words, let
\begin{eqnarray} 0 =\gamma^a(\partial_a - p_a)\psi,\end{eqnarray}
which in turn gives the fundamental wave equation
\begin{eqnarray} 
\gamma^a\partial_a\psi &=& \gamma^ap_a \psi.
\end{eqnarray}
This immediately suggests the particular solution  
\begin{eqnarray}\psi=e^{\int^x p^ady_a}\xi(p_0).\end{eqnarray}
Moreover, if the gauge term describes a test particle of mass $m$
moving along a timelike geodesic as defined in (8), then 
\begin{eqnarray} \gamma^a\partial_a\psi = mc\psi
.\end{eqnarray}
Once again, we have obtained a Dirac equation. 
\newline

To conclude this section, we summarize the results with the following theorem:
\begin{thm} Let $\xi(p)=[\frac{d}{ds}\psi^i(s)]e_i$, where $mcs=\int^xp^adx_a$
along a timelike geodesic then $\gamma^ap_a\xi=mc\xi$  iff
$\psi(p)=\psi^i(\int^xp^adx_a)e_i$ is a solution of
$$\frac{\partial }{\partial s}\psi(p)=\gamma^a\partial_a\psi(p),$$
where $\int^xp^adx_a$ is Lorentz invariant, 
and integration is taken along the curve
with tangent vector $p^a=m\frac{dx^a}{d\tau}$, where $\tau$ is proper time.
\end{thm}
{\bf Proof:} Noting that $\psi(p)=\psi^i(\int^x p^adx_a)e_i$ and assuming
$\gamma^ap_a\xi=mc\xi$ then
\begin{eqnarray} \gamma^a\frac{\partial \psi}{\partial x^a}&=&
\gamma^ap_a\frac{\partial\psi^i}{\partial s}e_i\\
&=&\gamma^ap_a\xi\\
&=&mc\xi \qquad \mbox{given}\\
&=&\frac{\partial \psi}{\partial s}
\end{eqnarray} To prove the converse it is sufficient to substitute
$\psi(p)=\psi^i(\int^xp^adx_a)e_i$ into $\tilde{\partial_s}\psi=
\partial_s \psi$ to get answer.

\begin{cor} In the case of $\psi^i(\int^xp^adx_a)=e^{\int^xp^adx_a}$ then 
$\gamma^a\partial_a\psi = mc\psi$.
\end{cor}
{\bf Proof:} Clearly $\frac{\partial \psi}{\partial s} = mc \psi$.
\begin{cor} If $\psi(\int^x p^adx_a)=f(\int^x p^adx_a)u$ where $u$ is a spinor 
independent of $x^a$ then the equation
\begin{eqnarray}\tilde{\partial_s}fu = 
\frac{\partial f}{\partial s}u.\end{eqnarray}
has the same solutions as 
\begin{eqnarray}\tilde{\partial_s}\psi = 
\frac{\partial }{\partial s}\psi.\end{eqnarray}
\end{cor}
{\bf Proof:} Substitute. 
\newline 

By way of conclusion,  note that if equation (8) is relaxed 
to incorporate gauge terms other than mass, such as electric charge,
then equation (6)
can be extended to incorporate the electromagnetic field and in particular,
the motion of the electron in a hydrogen atom.
We will discuss this in more
detail later. For the moment suffice it to say that if
$p_a=p^{\prime}_a-\frac{e}{c}A_a$ with $p^{\prime}_a$ corresponding to the
mass gauge component and subject to the conditions
of equation (8) and $A_a$ corresponding to the electromagnetic potential, then 
\begin{eqnarray} \gamma^a\left(\partial_a +\frac{e}{c}A_a\right)\psi 
= (mc+\frac{e}{c}A_0) \psi.\end{eqnarray} 
\section {Covariance}

Theorem 1 also enables us to write down a covariant form for
the generalized Dirac equation which depends directly upon the covariance of 
its dual metric equation.
We begin by showing that 
the equation $\tilde{ds}\xi=ds\xi$ is covariant under Lorentz transformations.
Specifically, if
$dx^a=\frac{\partial x^a}{\partial x^{\prime b}}dx^{\prime b}
=\Lambda^a_b dx^{\prime b}$ then
$\tilde{ds}\xi=ds\xi$ transforms under Lorentz transfomations 
$D(\Lambda(x))=D(x)$ into
\begin{eqnarray}D(x)\tilde{ds}\xi(x) = ds D(x)\xi(x).\end{eqnarray}
Now the left hand side can be rewritten as
\begin{eqnarray} D(x)\tilde{ds}\xi&=& D(x)\tilde{ds}D^{-1}(x)D(x)\xi(x)\\
&=&\tilde{ds^{\prime}}D(x)\xi(x),\qquad\mbox{where}\ \tilde{ds^\prime}\equiv
\gamma_adx^{{\prime}a}\\
&=&\tilde{ds^{\prime}}\xi^{\prime}(x^{\prime}).
\end{eqnarray}
Equating the two equations (27) and (30) then gives
\begin{eqnarray} \tilde{ds^{\prime}}\xi^{\prime}(x^{\prime})=
ds\xi^{\prime}(x^{\prime}),
\end{eqnarray}
which establishes the covariance. 

To show the covariance of the corresponding Dirac equation, we re-write
equation (6) in the form
\begin{eqnarray}(\tilde{\partial}_s\psi^i)e_i=
\frac{\partial \psi^i}{\partial s}e_i.\end{eqnarray}
From Theorem 1, we already know that this is equivalent to the the covariant
metric $\tilde{ds}\xi=ds\xi$ provided $\xi=(d\psi^i/ds)e_i$, where $s$ 
indicates differention along timelike geodesic parametrized by $s$ (recall
$mcs=\int^xp_adx^a$).
It follows that $\tilde{ds}\xi=ds\xi$ is covariant iff $\gamma_ap^a\xi=mc\xi$ is
covariant iff $\gamma_ap^a(d\psi^i/ds)e_i=mc(d\psi^i/ds)e_i$ is
covariant iff equation (32) is covariant with respect to the Lorentz 
transformation $D(x)$, along the geodesic. Moreover, this latter restriction 
of motion along a geodesic, may actually be relaxed and the following more
general theoren can be proven:

\begin{thm} The Dirac equation defined over the manifold $({\cal M},g)$ is
Lorentz covariant under the transformation $D(x)$ defined with respect to a
tetrad $e_i(x)$, provided the equation is written in the form
$$\gamma^a\frac{\partial \psi^i}{\partial x^a}e_i(x)=
\frac{\partial \psi^i}{\partial s}e_i(x).$$
\end{thm}
{\bf Proof:} Let $D(x)$ be a local Lorentz transformation at $x$ then: 
\begin{eqnarray} D(x)\tilde{\partial_s}\psi(s)&=&
D(x)(\tilde{\partial_s}\psi^i)e_i\\  
&=&(D(x)\tilde{\partial_s}\psi^i)D^{-1}(x)D(x)e_i\\
&=&(\tilde{\partial_{s^{\prime}}}\psi^i)e^{\prime}_i\\  
&=&\left(\frac{\partial\psi^i}{\partial s}\right)e^{\prime}_i 
\end{eqnarray}  
\newline

\noindent Remark: In regular Minkowski space, the covariant form of the 
generalized Dirac equation can be reduced to the form of equation (6).
\section {\bf Wave Equations for Geodesics}

At this stage the reader may be wondering how the usual formulation of quantum 
mechanics emerges. Indeed, the wave equations above seem to express the
wave equations of classical mechanics more than quantum mechanics, in that
there is no expression for Planck's constant $h$, 
nor does the the expression $i=\sqrt{-1}$ appear with the 
operators. With regard to the latter point, we note that $i$ could be seen
as absorbed into the $\gamma$ matrices, but we postpone a full discussion of 
this until the next section.  
First, we  analyze the solutions of the wave
equation for a massless particle from three perspectives 
to help us better grasp the formal difference
between classical and quantum  mechanics. 
Later on, we will formulate the axioms of 
quantum mechanics as suggested by our analysis. 

The linearized metric for a massless particle is given by   
\begin{equation}0=\gamma^0 cdt-\gamma^1dx_1-\gamma^2dx_2 -\gamma^3dx_3
\end{equation} from which
it follows by the canonical correspondence established above that
the associated wave equation for the particle is given by:
\begin{equation}0 =\gamma_0\frac {\partial \psi}{c\partial t}-\gamma_1\frac
{\partial \psi}{\partial x_1}-\gamma_2\frac {\partial
\psi}{\partial x_2}-\gamma_3\frac {\partial \psi}{\partial
x_3}.\end{equation}
This is the Dirac equation for a massless particle. 
Squaring this out we get the Klein-Gordan
equation for a massless particle :
\begin{equation} \frac {1}{c^2}\frac {\partial^2 \psi }{\partial
t^2}=\sum_{i=1}^3 \frac {\partial^2 \psi}{\partial x^2_i}.
\end{equation} 

Note that, in this formulation, solutions of the massless Dirac
equation
are also solutions of the usual massless Klein-Gordan equation, although
as we shall see later in the potential well problem (section 7),
eigenfunction solutions of
the Klein-Gordan equation are not necessarily eigenfunctions of the generalized
Dirac equation. In this regard, it should also be noted that the Klein-Gordan 
equation simply prescinds from any discussion of spin or equivalently, it may be
considered as the equation for a spin 0 particle. In contrast
the Dirac equation has non-zero spin value solutions.

This also raises the question of quantum statistics. It has been noted in a 
previous paper \cite{poh} 
that Fermi-Dirac statistics is a consequence of indistinguishable particles
forming spin-singlet states, 
while Bose-Einstein statistics follows as a consequence of breaking the 
rotational invariance associated with the singlet states. Moreover,
the easiest way for this breaking to occur is for the spin states of the 
particles
to be statistically independent. 
It follows as a trivial consequence of the above theory
that bosons cannot be second quantized as fermions and
fermions cannot be second quantized as bosons, in that
particles which are forming spin-singlet states 
with probability one cannot
be considered statistically independent. It also follows that spin 0 particles
must obey Bose-Einstein statistics. For example,  if $S$ and $T$ represent the 
spin observables of two particles such that $P(S=0)=P(T=0)=1$ then 
$P(S=0, T=0)=P(S=0)P(T=0)=1.1=1$, and hence the spin observables $S$ and $T$
are statistically independent.

However, it
is possible to make and break the spin-singlets. Transposed into the
context of quantum field theory this means that the wave function
of singlet state particles will have the anti-commutator equal zero 
while the wave function for 
statistically dependent particles will have the commutator equal zero. 
This result clearly differs from the usual form of the spin-statistics theorem,
in that the key to understanding the above version of the theorem lies in the 
rotational invariance of the singlet states, while in the conventional form
the theorem associates the different types of statistics with spin values.   
Nevertheless, in spite of the difference in interpretation and approaches,
algebraically the two approaches are easy to reconcile. 
In Pauli's version of the
theorem, the angular momentum operators $L_i$, in the case of fermions, 
obey the Lie Algebra commutator relationships
$[L_i,L_j]=i\epsilon_{ijk}L_k$ and the anti-commutator relationships
$\{L_i,L_j\}=0$. In the version mentioned above, the spin operators $S$ and $T$ 
associated with the two distinct particles respectively in a
spin-singlet state, are rescaled by defining $S=T=nL$, where $n$ is an
integer, and obey the commutator
relationships $[S_i,T_j]=[S_i,S_j]=in\epsilon_{ijk}S_k$. This rescaling
allows us to distinguish, for example,  spin 1/2 particles from spin 1 
particles by taking
n=1 and n=2 respectively in the above relationships.
In both cases, the statistics is determined by the commutator
relationships, and if in addition, we permit only $n=1$ Pauli's result 
necessarily follows. However, if we
permit different values of $n$ then spin-value ceases to be dependent
on commutator relationships, and the generalized Dirac Algebra needs to
be interpreted differently. 

In this context, rotational 
invariance
offers an alternative interpretation of the underlying Dirac Algebra, both
in Pauli's original version of the theorem and the current formulation.
For example, in the case of two body ``bosonic'' 
composites of spin 0 built out of spin-singlet states, as in the BCS-Anderson
theory of superconductivity, the statistics can be explained by partially 
relaxing
the indistinguishability condition. Specifically, n-Cooper pairs can be
viewed as obeying the statistics of n-indistinguisable spin-singlets of spin 0,
as distinct from the statistics of 2n-indistinguishable particles forming
n spin-singlets, where in this latter case, particles can switch from
one singlet to another. Finally, it is also worth noting that this revised
version of the spin-statistics theorem is strongly supported by an
argument based on Clebsch-Gordan coefficients.
The interested
reader is referred to the reference above for more details. We now return to
our previous discussion.

Since the wave equation emerges from the structure of space-time
itself, the question arises as to how to distinguish classical
mechanics from quantum mechanics. We investigate this by
analyzing
the motion of a massless particle in a Minkowski space,
subject to different sets of boundary conditions. In the first
case we consider the motion of a classical massless particle moving 
on the x-axis with
uniform velocity $c$, but constrained by two mirrors placed at
$x=0$ and $x=\xi$ to move uniformly on the interval $[0, \xi]$.
We
will assume that perfect reflection takes place at the mirrors
and
that no energy is exchanged.  In this case, the equation of motion for
a strictly classical particle with position $x=0$ at $t=0$ is given by:
\[ x=\left\{\begin{array}{ll} ct - 2n\xi, &\mbox 
{for $t\in [\frac {2n\xi}{c}, \frac
{(2n+1)\xi}{c}]$}\\
2(n+1)\xi -ct &\mbox {for $t\in [\frac {(2n+1)\xi}{c}, \frac
{(2n+2)\xi}{c}],$}
\end{array}\right. \]
and its wave function $\psi (x,t)$ takes on the form
\[ \psi(x,t)=\left\{\begin{array}{ll} \delta [k(x-ct)] &\mbox{for $\qquad x-
ct=-2n\xi$}\\
\delta[k(x+ct)] &\mbox{for $\qquad x+ct=2(n+1)\xi$}\\
0 &\mbox{otherwise.}
\end{array}\right. \]
The wave function in this case pinpoints the position of the
particle with probability 1. Moreover, there is no restriction on
the energy (implicit in the term $k$) in this case. 
Theoretically, it may have
values ranging from 0 to $\infty$.

However, the classical particle is an idealized situation.  In
reality, the position of a massless particle constrained to move on the
line
is unknown and any attempt to know its exact position will be
subject to Heisenberg's uncertainty relations, which we will formulate
in the next section. In other words, its
exact position can not be known in principle, because any attempt
to pinpoint it will scuttle the position and defeat the whole
purpose of the experiment.  The best we can do is to
describe the position by means of a uniform probability density
$f(x-ct)=1/\xi$ for $x \in [0, \xi]$ which also suggests writing
$\psi (x,t)=
e^{\pm ik(x-ct)}/\sqrt \xi$ to preserve both the boundedness and the periodic
motion of the particle, as described by the above wave equation.  
This does not mean that causality is
violated nor that the particle does not have an exact position, at least in the
above case.
It simply affirms that our initial conditions have to be defined
statistically, and also in such a way as to reflect the periodic motion
of the particle. As a consequence the future evolution of the wave
function of the
system is best interpretated in a statistical way. Finally, note
that in this model the energy of the particle can once again vary
from 0 to $\infty$ in a continuous manner.

Thirdly, the particle may be constrained to move in a potential
well in such a way that the wave function is continuous (= 0) at
the boundaries. In the case of the above problem, this means
that the wave function has harmonic solutions of the form
$\psi(\xi,t)=Ae^{i\nu^{\prime} t}\sin(kx)$, 
where $A$ is a
constant. Substituting, we will find that
$k=\frac{n\pi}{\xi}$ and the photon energy becomes quantized and of the form
$E\equiv kc=\nu^{\prime}$. 
It should be noted that this solution corresponds to the 
motion of a harmonic oscillator, and is the key to the quantization process 
associated with quantum field theory in general\cite{mil}. Indeed, if we
were to rescale our units of energy by defining $\nu^{\prime}=h\nu$, then
$E=h\nu$, with $h$ having units $J.s$, and the standard wavelength becomes 
$\lambda=\frac{ch}{E}$. In the next 
section, $h$ will be introduced in a more formal way.

The purpose of the above three examples is to highlight the
importance of the boundary conditions when distinguishing between
a classical type problem and a quantum mechanical problem, a
point also stressed by Lindsey and Margenau \cite{lin}.   
Classical and quantum laws are not in opposition to each other.
There is not one set of laws on the microscopic level and another
on the macroscopic.  On the contrary, classical and statistical
methodologies are complimentary to each other and are in
principle, applicable at all levels. However, on the microscopic
level, statistical fluctuations will be more pronounced 
and consequently in practice
(and in principle) the effects
associated with quantum physics will become more
apparent.
\section {Quantum Mechanics and Hilbert Spaces}

The above analysis permits us to better understand something of the difference
between quantum mechanics and classical mechanics from the perspective 
of general relativity.
As we have noted, it suggests the difference is to be found in the 
boundary conditions,
which in the case of quantum mechanics is subjected to statistical conditions. 
With this in mind, we formulate a few axioms which not only
respect the manifold
structure of general relativity, but also enable us to distinguish quantum
mechanics from classical physics, in a formal way.

Essentially what we 
have noted is that the metric of general relativity forces (real) 
eigenvalue solutions for the
free particle of the form
$$\frac{\partial\psi(\int p^adx_a)}{\partial x_a}=p^a\psi(\int p^adx_a).$$
However, since the choice of eigenfunction associated with the  specific
eigenvalue in this case is not unique, we restrict ourselves for the purpose
of quantum mechanics to those eigenfunctions $\psi(t,{\bf x})$ such that
for each $t$,
$\psi(t,{\bf x})\in L^2(E^3)\times H$, where $H$ is a 4-dimensional Hilbert 
space. Also, we associate the dual of the 1-form $dx$
with the self-adjoint partial differential operator $i\hbar\partial/\partial x$,
where $\hbar$ is a constant. Consequently by defining the dual in this way, we 
not only find that 
\begin{eqnarray}dx(i\hbar\partial_x)\equiv i\hbar\partial_xx=i\hbar ,
\end{eqnarray}
but it can also be linked to the uncertainty principle (see below). 
At first, this may seem artificial but actually if we look more closely 
at equation (6)
we will find that to associate the operator $-i\partial_x$ with a real
valued momentum eigenvalue is already implicit it this equation, and indeed is 
a consequence of the signature of the metric tensor 
$\eta_{ab}=\{\gamma_a,\gamma_b\}$. In particular, if we let  
$\gamma_a=i\alpha_a$ for $a=1,2,3$, and set 
$\gamma_0=\alpha_0$, then the $\alpha_a$'s are the generators of the Dirac
Algebra $SL(2,C)$. 
Also $\gamma_ap^a=\alpha_a(ip^a)$ and the linearized metric (3) can be 
written in the explicit form 
\begin{eqnarray}\tilde{ds} = \alpha_0dx^0+i\alpha_1dx^1+i\alpha_2dx^2+
i\alpha_3dx^3,
\end{eqnarray}
which in order to maintain the invariant relationship 
$\left<\tilde{ds},\tilde{\partial}_s\right> =1$ (cf eqn. (5)), gives
\begin{eqnarray}\tilde{\partial}_s = \alpha^0\partial_0-i\alpha^1\partial_1-
i\alpha^2\partial_2-i\alpha^3\partial_3.
\end{eqnarray}
In other words, if we let $\alpha_a$ obey the Dirac Algebra, then we can
associate the momentum operator with $-i\partial_a$ in a natural way, with 
eigenvalues $-p_a$ for $a=1,2,3$, where $p_a=dx^a/d\tau$ for
each $a$. Finally, let $\hbar=h/2\pi$ where $h$ is Plancks
constant  and rescale the momentum operator
by writing $-i\hbar\partial_a$ in place of $-i\partial_a$ to obtain the
usual form of quantum mechanics. 

This too
may seem artificial, but in reality we are free to choose any scale we wish.
This being the case, we choose $\hbar$ because it seems to be the scaling 
constant,
which nature uses. Moreover if we multiply across by $-i\hbar\alpha^0$ and 
note that
$\hat{\alpha}_a\equiv -i\alpha_0\alpha_a$ obeys the same Dirac Algebra as 
$\alpha_a$, then
the Dirac equation (6) can be rewritten as:
\begin{eqnarray}
-i\hbar{\alpha^0}\frac{\partial\psi}{\partial s}=(-i\hbar\partial_0-i\hbar
\hat{\alpha}^1\partial_1-i\hbar
\hat{\alpha}^2\partial_2-i\hbar\hat{\alpha}^3\partial_3)\psi.\end{eqnarray}
In particular, if we denote the eigenvalue of $i\hbar\partial_0$ associated 
with
the eigenfunction $\exp(-i\int^x p_adx^a)$ by $p_0=-iE/(\hbar c)$ 
then Cor. 1 gives
\begin{eqnarray}
\hat{\alpha}^0mc^2\psi=(E-ic\hbar
\hat{\alpha}^1\partial_1-ic\hbar
\hat{\alpha}^2\partial_2-ic\hbar\hat{\alpha}^3\partial_3)\psi\end{eqnarray}
which gives us back the usual form of the Dirac equation. Note too that $\hbar$
has been absorbed into the energy terms.

Finally based on the above discussion, we formulate a few axioms which not only
respect the manifold
structure of general relativity, but also enable us to distinguish quantum
mechanics from classical physics, in a formal way.

\begin{Def} Space-time is a four dimensional manifold $({\cal M},g)$.
\end{Def}

\begin{Def} At every point $p$ of $\cal M$ there is a tangent
vector space $T_p(\cal M)$ with tetrad basis  
$\{-i\hbar \partial_1,-i\hbar  \partial_2,
-i\hbar \partial_3, -i\hbar \partial_4\}$ and a dual 1-form space, 
denoted by $T^*_p$ with
basis $\{dx_1,dx_2,dx_3,dx_4\}$ at $p$, defined by 
$dx^a(i\hbar\partial_b)=
i\hbar\frac{\partial x^a}{\partial x^b}=i\hbar\delta^a_b$.
\end{Def}

\begin{Def} Quantum mechanical operators are elements of $SL(2,\cal C)$ 
Dirac Algebra, which can be viewed as a representation of the vector spaces 
$T_p$ and $T^*_p$ on $\cal M$.
\end{Def}

\begin{Def} Each element of $SL(2,\cal C)$ algebra acts on the Hilbert Space
$L^2(E^4)\times H$, where $H$ is a 4-dimensional Hilbert Space.The elements
$\psi\in L^2(E^4)\times H$ are called the states of the system.
\end{Def}

\noindent Remark: It follows from the definition of the Hilbert Space that if 
$\psi\in L^2(E^4)\times H$ then for each $t$,
$\psi(t,{\bf x})\equiv \psi_t({\bf x}) = \psi_t^ie_i \in L^2(E^3)\times H$, 
where each $e_i \in H$  and an inner product exists
such that $\left<\psi_t, \psi_t\right>=\int (\psi^*)^i\psi_id^3x$.
Moreover, if $\psi_t({\bf x})$ is normalized for each $t$ then 
$\psi^*\psi$ can be interpreted as a probability density function for
position.

\begin{lem} Let $d{\tilde f}$ and ${\tilde \partial_x}$ be the $SL(2,\cal C)$
representation of $df\in T^*$ and $\partial_x \in T$ respectively, then
$\left<d{\tilde f},{\tilde \partial_x}\right>\psi=
[{\tilde \partial_x}, {\tilde f}]\psi$, where $\psi\in L^2(E^4).$
\end{lem}
{\bf Proof:} Note $d{\tilde f}=\frac{\partial f}{\partial x^a}d{\tilde x}^a$. 
Therefore
$$\left<d{\tilde f}, 
{\tilde\partial}_a\right>\psi=\frac{\partial f}{\partial x^a}\psi=
[\partial_a, f]\psi.$$
The Lemma has been proven.

\begin{lem} (The uncertainty relationships)
Let ${\tilde X}=\int^{x(s)}_0d{\tilde X}$ and 
${\tilde P}=i\hbar{\tilde \partial_s}$ 
be the $SL(2,\cal C)$
representations of position and momentum
respectively, defined along a curve of length s. Also
let ${\bar X}\equiv \int \psi^*{\tilde X}\psi ds$, 
${\bar P}\equiv \int \psi^*{\tilde P}\psi ds$, 
$\Delta^2 {\tilde X}\equiv \int \psi^*({\tilde X}-{\bar X})^2\psi ds$ 
and $\Delta^2 {\tilde P}\equiv \int \psi^*({\tilde P}-{\bar P})^2\psi ds$ 
then 
$\Delta {\tilde X}\Delta {\tilde P}\ge \frac{\hbar }{2}\left|\int^s
\psi^*\frac{1}{i\hbar}({\tilde P}{\tilde X}-{\tilde Q}{\tilde X})\psi ds\right|$. In particular, in the case of the components ${\tilde X}^a, {\tilde P}^a$ 
we get $\Delta {\tilde X}^a\Delta {\tilde P}^a\ge \frac{\hbar }{2}$.
\end{lem}
{\bf Proof:} Usual proof using Cauchy-Schwartz inequality.
\section{ Classical Mechanics}

The above formulation lays the ground work for distinguishing 
classical from
quantum mechanics. Indeed, we have seen that quantum theory is highly dependent 
upon $\hbar>0$ and states $\psi\in L^2(E^4)$. Moreover, 
this suggests that classical mechanics can be obtained by relaxing one 
of these two conditions either by letting
$\hbar \to 0$ or by choosing $\psi \notin L^2(E^4)$ or both. 
In principle what distinguishes a quantum particle from a classical one is
that in contrast to quantum mechanics, 
the position and momentum of a classical particle
can be fully pinpointed and localized. 
For example, if we reconsider the case   
of a particle moving uniformly between two mirrors, where the 
initial position is unknown, then as has already been noted in section 4, 
the wave function is given by 
$exp(ikx)/\sqrt{\xi}$.
However, if this really were a classical situation then in principle both
the position and momentum of the particle could 
be localized and measured exactly,
as the distance between the mirrors $\xi$ shrinks to 0. Moreover, the resulting 
wave function at 
any time $t$ would be of the form $\psi(0)=\sum_ka_k\delta^{(k)}$. 
This last equation follows from a well 
known result in distribution theory \cite{duf}, which states
\begin{thm} A distribution $T$ which has a support of one point (i.e., is
equal to zero except at one point) is a finite linear combination of the
Dirac function and its derivatives: $T=\sum_ka_k\delta^{(k)}$.
\end{thm}
Based on this we can formally state that
\begin{Def} A classical particle is a particle whose position operator $T$
at any time t has support of one point. 
\end{Def}  
It follows from this definition that the momentum operator $\tilde P$ has 
a support of one point. It also follows from the definition and
Theorem 1 that for a classical particle 
with constant momentum situated at
$(x_0)$ on a geodesic, the wave function is given by 
$\delta^4(p^a(x-x_0)_a)$. Also the set of operators $T$ in definition 5,
clearly form a subspace of the solution set of the Dirac equation. Indeed,
the set $\{\delta, ... , \delta^{(k)} ...\}$ is a spanning set for 
this subspace.
Hence, the  uncertainty in observing 
its value is 0 everywhere except
at the single point, and the standard deviation
$\Delta{\tilde X}$ and $\Delta {\tilde P}$ are also zero. 
In other words, the uncertainty
principle fails for a classical particle.
Moreover, in order for nature to circumvent classical solutions, it is
sufficient that there be a fundamental unit of wavelength given by 
$\lambda =\frac{ch}{E}=\frac{2\pi c\hbar}{E}$,
such that $\lambda>0$ whenever $\hbar>0$, which is another way of saying that
if $\hbar>0$ then classical solutions need not exist. It also strongly
suggests that the process of localizing a particle for measurement is 
equivalent to confining (at least during the measuring 
process) the particle to a box. This, in turn, hints that
in terms of wave-particle duality, particle properties emerge
when we attempt to experimentally localize and isolate the wave, causing a
discontinuity in the quantum solutions, closely approximated by delta type
functions. We have seen an example of this
above, when we considered a particle moving
uniformely between two mirrors with wave-function $e^{\pm ik(x-ct)}/\sqrt{\xi}$.

In concluding this section, we note that classical solutions to the generalized
Dirac Equation
describing the motion of a particle, can be reduced to distribtions
corresponding to point masses as described in Theorem 3 above, 
and live on a larger space than the
$L^2$ functions associated with quantum mechanics. To better understand
this point, it might be useful to recall the definition of $L^p$ spaces, and
some of their properties\cite{fol}.
Consider a fixed measure space $(X, {\cal M}, \mu)$. Let $f$ be a measurable
function on $X$ such that
\begin{eqnarray}\|f\|_p=\left(\int|f|^pd\mu\right)^{\frac 1p}
\end{eqnarray}
then we define
\begin{Def} $L^p(X,{\cal M}, \mu)=\{f:X\to {\cal C}: \|f\|_p<\infty\}.$
\end{Def} 
Also, in general, we can define a bounded linear functional on $L^p$ by
$\phi_g(f)=\int fg$, such that $g\in L^q$ where $1/p + 1/q =1$, and
$\phi_g \in (L^p)^*$, the dual space of
$L^p$.
In the case of $p=2$, the $L^2$ space is also a Hilbert space and therefore,
its own dual. In turn this allows us to formulate quantum mechanics in a
very elegant and simple manner. However, it the case of classical
solutions, as described above, the Dirac $\delta$ functional is usually
interpreted as a functional on the set of continuous functions of 
compact support denoted by $C^{\infty}_c(X)$, which in turn are dense in $L^p$, where
$1\le p < \infty$.  
Moreover, in the case of
a finite (probability) measure, $\L^p\subset \L^1$, for $p\ge 1$.
With this distinction in mind, it now follows from our 
formulation that general relativity while pemitting a natural unification
of both quantum and classical mechanics by means of the generalized Dirac
Equation, also permits a distinction by means of
$L^2$ functions and distributions which are duals
of $L^1$ functions. 
\section {One dimensional Potential Well}

Consider a quantum particle moving in a Minkowski space such that there is 
zero potential between $(-a,a)$ and constant potential $V$ for $|x|\ge a$.
Then the generalized Dirac equation will in both cases reduce to
\begin{equation} \alpha^0\hbar\frac{\partial \psi}{\partial ct}-i\hbar
{\alpha^1}
\frac{\partial \psi}{\partial x}=\hbar\frac{\partial \psi}{\partial s}.
\end{equation}
Clearly wave function solutions of the form
\[\psi =\left\{\begin{array}{ll} Ae^{\hbar^{-1}[Et-px]}+Be^{-\hbar^{-1}[Et-px]}\qquad &x\le -a\\
 Ce^{i\hbar^{-1}[Et-px]}+De^{-i\hbar^{-1}[Et-px]}\qquad &|x|<a\\
 Fe^{\hbar^{-1}[Et-px]}+Ge^{-\hbar^{-1}[Et-px]}\qquad &x\ge a
\end{array}
\right. \]
can be found, although $\psi$ is not necessarily an eigenfunction of
equation (46). Note the $s$ dependency follows from the
Lorentz invariant relationship $mcs=Et-px$,
and that in order to satisfy the boundary counditions at
$\pm \infty$, $A=G=0$. Also smooth continuity conditions can be
imposed at $\pm a$ to fully solve the system in the usual way.

On the other hand, if we begin with the conventional form of the Dirac
equation as given in Cor 2, namely
\begin{equation} \alpha^0\hbar\frac{\partial \psi}{\partial ct}-i\hbar
{\alpha^1}
\frac{\partial \psi}{\partial x}=mc\psi
\end{equation}
then solutons of the form
$\psi = Ce^{i\hbar^{-1}[Et-px]}+De^{-i\hbar^{-1}[Et-px]}$
cannot exist if both $C$ and $D$ are non-zero, as is the case in the
conventional potential well problem, where $C=\pm D$. 
In other words, equation (47) cannot be used to describe a potential well 
problem.
It is also worth noting that if one were to square out either (46) or
(47) then both of these equations would have $\psi$ as a permissible
solution.
This ambiguity arises from the non uniqueness of 
the square root, especially of the $\alpha^1$ matrix which can be defined to
be either
$$\left(\begin{array}{cc}
0 & \sigma_1\\
\sigma_1 & 0
\end{array}\right)
\qquad \mbox {or}\qquad
\left(\begin{array}{cc}
0 & \sigma_2\\
\sigma_2 & 0
\end{array}\right)$$
where $$\sigma_1=
\left(\begin{array}{cc}
0 & 1\\
1 & 0
\end{array}\right)\qquad \mbox {and}\qquad \sigma_2=
\left(\begin{array}{cc}
0 & i\\
-i & 0
\end{array}\right)$$
but not both simultaneously.
\section {Application to  the Schwarzschild metric}

We directly apply the above theory to a test particle of mass $m$ moving along
a timelike curve parametrized by $\tau$
in a Schwarzschild space, whose metric is defined by \cite{wein}:
$$ds^2=B(r)dt^2 - A(r) dr^2 - r^2 d{\theta^2} - r^2\sin^2\theta d{\phi^2}.$$
This allows us to define tetrad coordinates 
$dx^0=B^{1/2}dt$, $dx^1=A^{1/2}dr$,
$dx^2=rd\theta $, $dx^3=r\sin(\theta)d\phi$, which on substituting into 
equation (8) gives 
\begin{eqnarray} ds\xi = [\gamma_0B^{\frac12}dt-\gamma_1A^{\frac12}dr
-\gamma_2rd\theta -\gamma_3r\sin(\theta)d\phi]\xi.
\end{eqnarray}
This is also equivalent to  
\begin{eqnarray} c\xi = [\gamma_0B^{\frac12}\dot{t}-\gamma_1A^{\frac12}\dot{r}
-\gamma_2r\dot{\theta} -\gamma_3r\sin(\theta)\dot{\phi}]\xi,
\end{eqnarray}
which can be seen as the equation of motion along a timelike geodesic.
The corresponding wave equation associated with the metric equation (48) 
is then given by
\begin{eqnarray}\frac{\partial \psi}{\partial s} =\gamma^0B^{-\frac12}(r)\frac{\partial \psi}
{\partial t}-
\gamma^1A^{-\frac12}(r)\frac{\partial \psi}{\partial r}-
\gamma^2\frac{1}{r}\frac{\partial \psi}{\partial\theta}
-\gamma^3\frac{1}{r\sin\theta}\frac{\partial \psi}{\partial \phi}.
\end{eqnarray}
To simplify the problem, we assume $\dot{\theta}=\dot{\phi}=0$. 
Hence, the equation of motion (49) reduces to
\begin{eqnarray}\gamma^0B^{\frac12}\dot{t}\xi-\gamma^1A^{\frac12}\dot{r}\xi=
c\xi \end{eqnarray}
where 
$$\gamma^0=\left(\begin{array}{cccc} 1 & 0 & 0 & 0\\
0 & 1 & 0 & 0\\
0 & 0 & -1 & 0\\
0 & 0 & 0 & -1
\end{array}\right)\qquad \mbox{and}\qquad
\gamma^1=\left(\begin{array}{cccc} 0 & 0 & 0 & i\\
0 & 0 & i & 0\\
0 & i & 0 & 0\\ 
i & 0 & 0 & 0
\end{array}\right).$$  
Multiplying out gives              
$$\begin{array}{rcc}
(B^{\frac12}\dot{t}-c)\xi_0-iA^{\frac12}\dot{r}\xi_3 &= & 0\\
(B^{\frac12}\dot{t}-c)\xi_1-iA^{\frac12}\dot{r}\xi_2 &= & 0\\
-(B^{\frac12}\dot{t}+c)\xi_2-iA^{\frac12}\dot{r}\xi_1 &= &0\\
-(B^{\frac12}\dot{t}+c)\xi_3-iA^{\frac12}\dot{r}\xi_0 &= &0
\end{array}$$
which can be solved to give
\begin{equation}(B^{\frac12}\dot{t}-c)\xi_0=iA^{\frac12}\dot{r}\xi_3\end{equation}
and
\begin{equation}(B^{\frac12}\dot{t}-c)\xi_1=iA^{\frac12}\dot{r}\xi_2.\end{equation}
Clearly many solutions are possible.  One solution is given by
$\xi_0=\xi_1=iA^{\frac12}\dot{r}$ and $\xi_3=\xi_2=(B^{\frac12}\dot{t}-c)$. 
Another
solution is given by $\xi_0=\xi_1=(B^{\frac12}\dot{t}+c)$ and $\xi_3=\xi_2=
-iA^{\frac12}\dot{r}$. Note that in this latter case, if we substitute into
equations (52) and (53) we obtain the equation of motion $B\dot{t}^2-A\dot{r}^2
=c^2$.
Also, recall 
from Theorem 1 that $\xi=\frac{\partial \psi^i}{\partial s}e_i$. Therefore,
\begin{equation}\psi^0=\psi^1=\int(B^{\frac12}\dot{t}+c)ds\end{equation}
and \begin{equation}\psi^3=\psi^2=-i\int(A^{\frac12}\dot{r})ds.\end{equation}

In the case of constant motion along a geodesic, the above reduces to 
$\psi = \psi(\int p_adx^a)=\psi(p_0x_0-p_1x_1)$ where $p_0=mB^{\frac12}\dot{t}$
and $p_1=mA^{\frac12}\dot{r}$ are constants of the motion. In particular, in
the case of motion inside a potential well of radius $x_1(r)\le l$, we find that
$C\exp[i(p_0x_0-p_1x_1)]+D\exp[i(p_0x_0+p_1x_1)]$ is a solution
of the squared out Dirac equation (expressed in tetrad coordinates)
\begin{equation} {\hbar}^2\frac{\partial^2\psi}{\partial s^2}={\hbar}^2
\frac{\partial^2\psi}{\partial x_0^2}-{\hbar}^2\frac{\partial^2 \psi}
{\partial x^2_1}. \end{equation}
Denoting the eigenvalues of 
$i\hbar \frac{\partial }{\partial s}$ by $mc$ and the eigenvalue of 
$i\hbar \frac{\partial }{\partial x_0}$ by $\frac{E}{c}$, this
equation reduces to
\begin{equation} (mc)^2\psi = \left(\frac{E}{c}\right)^2\psi+\hbar^2
\frac{\partial^2\psi}
{\partial x_1^2}.\end{equation}
Imposing the boundary conditions $\psi(x_0, x_1(r)=l)=0$ gives
\begin{equation}\psi=\exp(i(E/c)x_0)\sin(p_1x_1)=\exp(i(E/c)B^{\frac12}t)
\sin(p_1x_1),\end{equation} with
$p_1\equiv mk_n=\frac{n\pi}{l}$. Therefore the energy levels for the motion of the
particle along a geodesic given by $\dot{\theta}=\dot{\phi}=0$
inside the potential well, can be calculated from the equation
\begin{equation} (mc)^2\psi=\left(\frac{E}{c}\right)^2\psi-(mk_n)^2\psi.
\end{equation}
In other words,
$$E^2=m^2c^2(c^2+k_n^2)=m^2c^2B\dot{t}^2, 
\qquad n\ \mbox{an integer}.$$
Also, if we permit $l\to 2mG/c^2$ (the Schwarzschild radius), in such a way
that $p_0=m(1-\frac{2Gm}{c^2r})^{\frac12}\dot{t}$ and $p_1=
m(1-\frac{2Gm}{c^2r})^{-\frac12}\dot{r}$ remain constant then we can
interpret the above energy levels as the energy levels within a black hole
provided we work with spacelike geodesics and not timelike ones.

Finally, to conclude this section, note that 
in the case of a photon the wave equation along a null geodesic is given by
\begin{eqnarray}0 =\gamma^0B^{-1/2}(r)\frac{\partial \psi}{\partial t}-
\gamma^1A^{-1/2}(r)\frac{\partial \psi}{\partial r}-
\gamma^2\frac{1}{r}\frac{\partial \psi}{\partial\theta}
-\gamma^3\frac{1}{r\sin\theta}\frac{\partial \psi}{\partial \phi}.
\end{eqnarray}
\section {The Hydrogen Atom}

We now apply the techniques of this paper to describe the
hydrogen atom. More specifically, we describe the motion of the
electron lying within the Reissner-Nordstrom metric [14] of the
proton of mass $m_p$.  Linearizing this metric gives:
\begin{equation} ds =i\alpha_1 \biggl (1-\frac {2Gm_p}{c^2r}+\frac
{Ge^2}{c^4r^2} \biggr )^{-\frac 12}dr +ir(\alpha_2 d\theta
+\alpha_3 \sin \theta d\phi )+\alpha_0\biggl (1- \frac
{2Gm_p}{c^2r}+\frac {Ge^2}{c^4r^2}\biggr )^{\frac 12}c dt.\end{equation}    
The corresponding wave equation becomes:
\begin{eqnarray}
\frac{\partial}{\partial s}\psi^{\prime}(r,t)=\biggl \{
-i\alpha_1 \biggl (1-\frac {2Gm_p}{c^2r} + \frac
{Ge^2}{c^4r^2}\biggr)^{\frac 12}\frac {\partial }{\partial r}
- \frac ir \biggl (\alpha_2 \frac {\partial}{\partial
\theta}\\
+\alpha_3 \frac {1}{\sin \theta}\frac {\partial
}{\partial \phi }\biggr )   
+\alpha_0\biggl (1-\frac {2Gm_p}{c^2r}+\frac
{Ge^2}{c^4r^2} \biggr )^{-\frac 12} \frac 1c \frac {\partial
}{\partial t} \biggr \}\psi^{\prime}(r,t). \end{eqnarray}  
We will find on substituting that
$\psi^{\prime}= \psi^{\prime}(\int^x p^{\prime}_adx^a)$ is a solution, 
where $p^{\prime}_a=
p_a+\frac{e}{c}A_a$ and $A_a$ represents the vector potential associated
with the electromagnetic field of the proton. 
In particular, if we seek solutions of the form\newline
$\psi^{\prime}=\exp({\frac{ie}{c\hbar}\int A_adx^a})\psi(\frac{i}{\hbar}
\int p_adx^a)$ in tetrad coordinates then this can be reduced to 
the equation
\begin{equation}\left[c\sum_{a=1}^3\hat{\alpha}_aP^{\prime}_a+\alpha_0
(mc^2+eV)\right]\psi = E \psi,
\end{equation}
where $\hat{\alpha}_a=-i\alpha_0\alpha_a$,  
$P^{\prime}_a=-i\hbar\partial_a + \frac{e}{c}A_a$, 
$m$ is the mass of the electron and
$V$ is the electrostatic potential in the rest frame of the proton.
Note that this can be written in the usual form of the equation for the 
hydrogen atom $H\psi = E\psi$, provided $H\equiv   
\left[c\sum_{a=1}^3\hat{\alpha}_aP^{\prime}_a+\alpha_0
(mc^2+eV)\right]$. However, there is also an important difference. In this
formulation there is the presence of a $\alpha_0 eV$, instead of the usual
$eV$, but as we will see below this has advantages.  

If we let ${\cal G}_a=\frac{\partial V}{\partial x^a}$, assume
$A_a=0$ for a=1,2 and 3, then the 
``square''  of this equation,  reduces to
\begin{eqnarray}(-c^2\hbar^2\sum^3_1\partial^2_a
+(mc^2+eV)^2+i\hbar ec\hat{\alpha}_a
\alpha_0{\cal G}_a)\psi=E^2\psi.\end{eqnarray} 
Noting that $-i\hbar\hat{\alpha_a}{\alpha_0}=\alpha_a$ which can also be
seen as generators of the Dirac Algebra, this can be rewritten as:
\begin{eqnarray}(-c^2\hbar^2\sum^3_1\partial^2_a+(mc^2+eV)^2-\hbar ec\alpha_a
{\cal G}_a)\psi=E^2\psi.\end{eqnarray} 
Contrast this with conventional relativistic quantum mechanics
 where we would have found
\begin{eqnarray}(-c^2\hbar^2\sum^3_1\partial^2_a+m^2c^4-i\hbar ec\alpha_a{\cal
G}_a)\psi=(E+eV)^2\psi.\end{eqnarray} 
Notice that the complex number coefficient associated with the electric moment
$\cal G$ in this latter equation has been replaced with a real coefficient
in equation (66). In other words, the spin electric moment
$\frac{\hbar e}{2imc}$, which is a complex number,
 can be replaced with a real spin term 
$\frac{\hbar}{2mc}$ thus removing the ambiguity normally associated with
electron spin in the hydrogen atom. For example, Lindsey and Margenau 
\cite{mar} warn us ``not to take the electron spin to literally", 
because of the presence of the imaginery term. Happily, we can say that
with the above approach the difficulty is resolved.
\newline\newline
\section {Conclusion}

In this paper we have attempted to unify general relativity and 
quantum mechanics by viewing any metric as a dual of a wave equation. We 
have noted that the resulting wave equation contains the
usual Dirac equation of quantum mechanics as a special case.
We have also noted that the difference between quantum and classical mechanics 
seems to lie
in boundary conditions, with quantization (as distinct from quantum theory)
emerging when the wave
function is confined to a finite domain with continuous boundary conditions,
and classical mechanics being the result of delta-function type solutions
for the wave equation.
Indeed, this particular approach also highlights the harmonic oscillator as
a natural starting model for quantum field theory, by confining the quantum
particle to a box. Of course, our use of the word ``confined'' differs from
the usual non-Abelian gauge theory usage. Here the word, confined means 
considering solutions of the generalized Dirac equation constrained by periodic
boundary conditions. Nevertheless it would not surprise me if in the future
the two different types of confinement end up being linked.   

Overall, our approach was able to duplicate the standard results of 
quantum mechanics but
in addition, we were able 
to remove the anomaly of an imaginery electric moment, when solving the
hydrogen atom problem. This result in itself, should be sufficient
to encourage further development. I would also hope that
further studies will be carried out on the relationship between
confinement problems in general and quantum boundary conditions.

\end{document}